\documentclass[journal,final]{IEEEtran}
\usepackage{cite}

\ifCLASSINFOpdf
   \usepackage[pdftex]{graphicx}
   \graphicspath{{../pdf/}{../jpeg/}}
   \DeclareGraphicsExtensions{.pdf,.jpeg,.png}
\else
   \usepackage[dvips]{graphicx}
   \graphicspath{{../eps/}}
   \DeclareGraphicsExtensions{.eps}
\fi
\usepackage[cmex10]{amsmath}

\usepackage{mathabx}
\usepackage{longtable}
\usepackage{booktabs}
\usepackage{algorithm}
\usepackage{algorithmic}
\begin{document}
%
\title{User Association with Maximizing Sum Energy Efficiency for Massive MIMO Enabled Heterogeneous Cellular Networks}

\author{\IEEEauthorblockN{Tianqing Zhou}\\
\IEEEauthorblockA{School of Information Science and Engineering, Southeast University, Nanjing 210096, China\\
Email: {zhoutian930}@163.com\\
}
}


%


\maketitle

\begin{abstract}
In this paper, we design an association scheme to maximize the sum energy efficiency for massive multiple-input and multiple-output (MIMO) enabled heterogeneous cellular networks (HCNs). Considering that the final formulated problem is in a sum-of-ratio form, we first need to transform it into a parametric nonfractional form, by which we can achieve its solution through a two-layer iterative algorithm. The outer layer searches the energy efficiency parameters and multipliers associated with signal-interference-plus-noise-ratio (SINR) constraints using Newton-like method, and the inner layer optimizes the association indices using Lagrange multiplier method. In fact, the inner layer doesn't need iterative steps when the SINR constraints are not involved in the original problem, and then the whole algorithm should be a one-layer iterative one. As for the two-layer iterative algorithm, we also give the corresponding convergence proof. Numerical results show that the proposed scheme significantly outperforms the existing one in system throughput and network energy efficiency. In addition, we also investigate the impacts of the number of massive antennas and the transmit power of each pico base station on these association performances.
\end{abstract}


%
\IEEEpeerreviewmaketitle

\section{Introduction}
Heterogeneous cellular networks (HCNs) integrate macrocells and other small cells such as microcells, picocells and femtocells, where the macrocells are mainly used to guarantee the basic coverage but the small cells are introduce to further improve spatial reuse and coverage, e.g., eliminate coverage holes and relieve hot spots \cite{1}. In order to further enhance area spectral efficiency, the massive multiple-input and multiple-output (MIMO) technology has been implemented at macro base stations (BSs), which simultaneously transmits some independent data streams to multiple users sharing the same resource via a large-scale antenna system \cite{2}.
\par
Although massive MIMO is a promising solution to enhance the spectrum efficiency, the circuit power consumption increases with the number of massive MIMO antennas. Evidently, energy-efficient radio resource management should be an important topic for massive MIMO systems and mentioned in the designs of some schemes.
\par
User association often assigns users to the different BSs available in the system, which is an indispensable element of radio resource management. So far, most researches on user association mainly refer to single antenna HCNs, such as \cite{3,4}, etc. The efforts in the literature toward user association in massive MIMO enabled HCNs are very few and still in infancy. In work \cite{5}, authors design some association schemes with various objectives including (achievable) rate maximization, proportional fairness and the user association with resource allocation. In work \cite{6}, authors consider an $\alpha$-utility maximization association for massive MIMO wireless networks. Instead of system throughput in works \cite{5,6}, authors in work \cite{2} take the energy efficiency as a key factor for the association design, and propose an energy-efficient association scheme to maximize the network-wide utility that is a function with respect to users' energy efficiencies.
\par
In this paper, we take account of an energy-efficient association from a novel perspective. Unlike work \cite{2}, we design an energy-efficient association scheme to maximize the sum energy efficiency under users' SINR (signal-interference-plus-noise-ratio) constrains. According to the form of the final formulated problem, it is easy to know that it is a sum-of-ratio maximization problem and hard to tackle. To solve it, we first need to transform it into a parametric nonfractional form. Then, we can develop a two-layer iterative algorithm with guaranteed convergence to achieve its solution. Specially, we search the energy efficiency parameters and multipliers associated with SINR constraints using Newton-like method in the outer layer, and achieve the association indices using Lagrange multiplier method in the inner layer. It is noteworthy that the inner layer doesn't need any iterative steps when the SINR constraints are not considered in the optimization problem, and then the whole algorithm becomes a one-layer iterative one. However, the SINR constraints are essential for guaranteeing the experience of users.

\section{SYSTEM MODEL}
Without loss of generality, we just take account of two-tier HCNs, where the first tier is consisting of macro BSs (MBSs) and the second tier is composed of pico BSs (PBSs). Significantly, MBSs and PBSs are also known as macrocells and picocells respectively. In such HCNs, the MBSs are fixed and deployed into a traditional cellular network, but the PBSs and users are uniformly and independently deployed at each macrocell. In addition, the MBSs implement large-scale antenna array but the PBSs just utilize single antenna \cite{2}.
\par
We assume that the set of BSs is $\mathcal{N}={{\mathcal{N}}_{m}}\cup {{\mathcal{N}}_{p}}$, where ${{\mathcal{N}}_{m}}$ denotes the set of MBSs and ${{\mathcal{N}}_{p}}$ represents the one of PBSs. We let the set of users be $\mathcal{K}$, and write the cardinalities of $\mathcal{N}$ and $\mathcal{K}$ as $N=\left| \mathcal{N} \right|$ and $K=\left| \mathcal{K} \right|$ respectively. Then, the SINR received by user $k$ from BS $n$ is
\begin{equation}\label{eq1}
{{\text{SINR}}_{nk}}=\left\{ \begin{split}
  & \frac{\kappa {{p}_{n}}{{g}_{nk}}}{{{\sum }_{j\in \mathcal{N}\backslash \left\{ n \right\}}}{{p}_{j}}{{g}_{jk}}+\sigma _{n}^{2}},\forall n\in {{\mathcal{N}}_{m}},\forall k\in \mathcal{K}, \\
 & \frac{{{p}_{n}}{{g}_{nk}}}{{{\sum }_{j\in \mathcal{N}\backslash \left\{ n \right\}}}{{p}_{j}}{{g}_{jk}}+\sigma _{n}^{2}},\forall n\in {{\mathcal{N}}_{p}},\forall k\in \mathcal{K}, \\
\end{split} \right.
\end{equation}
where ${{p}_{n}}$ represents the transmit power of BS $n$; ${{g}_{nk}}$ denotes the channel gain between BS $n$ and user $k$; $\sigma _{n}^{2}$ is the noise power of BS $n$; $\kappa ={\left( M-S+1 \right)}/{S}$; $M$ represents the number of MBS antennas; $S$ is the maximal number of downlink data streams that one MBS can simultaneously transmit on some given resource block (RB) with equal power allocation.
\par
Then, the achievable rate received by user $k$ from BS $n$ is
\begin{equation}\label{eq2}
{{r}_{nk}}=\left\{ \begin{split}
  & S{{\log }_{2}}\left( 1+{{\text{SINR}}_{nk}} \right),\forall n\in {{\mathcal{N}}_{m}},\forall k\in \mathcal{K}, \\
 & {{\log }_{2}}\left( 1+{{\text{SINR}}_{nk}} \right),\forall n\in {{\mathcal{N}}_{p}},\forall k\in \mathcal{K}, \\
\end{split} \right.
\end{equation}
\section{Problem Formulation}
To design an energy-efficient association for massive MIMO enabled HCNs, we first need to give the definition of energy efficiency. Specially, it is defined as the ratio of user's long-term (effective) rate to the power consumed by some associated BS. Finally, the energy-efficient association scheme is formulated as a sum energy efficiency maximization problem, and given by
\begin{equation}\label{eq3}
\begin{split}
  \underset{\boldsymbol{x}}{\mathop{\max }}\,&\ \sum\limits_{n\in \mathcal{N}}{\sum\limits_{k\in \mathcal{K}}{\frac{{{x}_{nk}}{{R}_{nk}}}{{{\rho }_{n}}{{p}_{n}}+p_{n}^{c}}}} \\
 & \text{s}\text{.t}\text{s.t. }\sum\limits_{n\in \mathcal{N}}{{{x}_{nk}}}=1,\forall k\in \mathcal{K}, \\
 &\sum\limits_{n\in \mathcal{N}}{{{x}_{nk}}\text{SIN}{{\text{R}}_{nk}}}\ge {{\tau }_{k}},\forall k\in \mathcal{K},\\
 &{{x}_{nk}}\in \left\{ 0,1 \right\},\forall n\in \mathcal{N},\forall k\in \mathcal{K}, \\
\end{split}
\end{equation}
where ${{x}_{nk}}$ represents the association index, and it is 1 if user $k$ is associated with BS $n$, 0 otherwise; ${{R}_{nk}}={{{r}_{nk}}}/{\sum\nolimits_{i\in \mathcal{K}}{{{x}_{ni}}}}\;$ denotes the long-term (effective) rate received by user $k$ from BS $n$; ${{\rho }_{n}}$ denotes the power amplifier coefficient of BS $n$; ${{\tau }_{k}}$ is the SINR threshold of user $k$; $p_{n}^{c}$ is the circuit power consumption of BS $n$ and given by
\begin{equation}\label{eq4}
p_{n}^{c}=\left\{ \begin{split}
  & \sum\nolimits_{i=0}^{3}{{{C}_{i0}}{{S}^{i}}}+M\sum\nolimits_{i=0}^{2}{{{C}_{i1}}{{S}^{i}}},n\in {{\mathcal{N}}_{m}}, \\
 & p_{n}^{c},n\in {{\mathcal{N}}_{p}}, \\
\end{split} \right.
\end{equation}
where ${{C}_{i0}}$ and ${{C}_{i1}}$ are the coefficients \cite{2}. In the formulated problem, the second constraint gives the minimal SINR requirements of associated users, and the first constraint show that some user can just be associated with only one BS.
\par
It is easy to know that the problem \eqref{eq3} is in a sum-of-ratio form. In addition, it is also a nonlinear and mixed-integer optimization problem. Evidently, it is very challenging for designers to develop effective algorithms. In the next section, we will try to solve it using the method proposed in work \cite{7}.
\subsection{Association Algorithm Design}
To solve the problem \eqref{eq3}, we need to transform it into a tractable form. To this end, we consider the following equivalent transformation:
\begin{equation}\label{eq5}
\begin{split}
  \underset{\boldsymbol{x},\boldsymbol{\omega }}{\mathop{\max }}\,&\ \sum\limits_{n\in \mathcal{N}}{\sum\limits_{k\in \mathcal{K}}{{{x}_{nk}}{{\omega }_{nk}}}} \\
\text{s.t. }& \sum\limits_{n\in \mathcal{N}}{{{x}_{nk}}}=1,\forall k\in \mathcal{K}, \\
 &\sum\limits_{n\in \mathcal{N}}{{{x}_{nk}}{{\text{SINR}}_{nk}}}\ge {{\tau }_{k}},\forall k\in \mathcal{K},\\
 &\frac{{{R}_{nk}}}{{{\rho }_{n}}{{p}_{n}}+p_{n}^{c}}\ge {{\omega }_{nk}},\forall n\in \mathcal{N},\forall k\in \mathcal{K}, \\
 &{{x}_{nk}}\in \left\{ 0,1 \right\},\forall n\in \mathcal{N},\forall k\in \mathcal{K}, \\
\end{split}
\end{equation}
To meet the demand of algorithm design (i.e., avoid ¡°0/0¡±), we relax the third constraint in the problem \eqref{eq5}, and thus have
\begin{equation}\label{eq6}
\begin{split}
  \underset{\boldsymbol{x},\boldsymbol{\omega }}{\mathop{\max }}\,&\ F\left( \boldsymbol{x},\boldsymbol{\omega } \right)=\sum\limits_{n\in \mathcal{N}}{\sum\limits_{k\in \mathcal{K}}{{{x}_{nk}}{{\omega }_{nk}}}} \\
\text{s.t. }&\sum\limits_{n\in \mathcal{N}}{{{x}_{nk}}}=1,\forall k\in \mathcal{K}, \\
 &\sum\limits_{n\in \mathcal{N}}{{{x}_{nk}}{{\text{SINR}}_{nk}}}\ge {{\tau }_{k}},\forall k\in \mathcal{K},\\
 &{{r}_{nk}}\ge {{\alpha }_{n}}{{\omega }_{nk}}\left( 1+\sum\limits_{i\in \mathcal{K}}{{{x}_{ni}}} \right),\forall n\in \mathcal{N},\forall k\in \mathcal{K}, \\
 &{{x}_{nk}}\in \left\{ 0,1 \right\},\forall n\in \mathcal{N},\forall k\in \mathcal{K}, \\
\end{split}
\end{equation}
where ${{\alpha }_{n}}={{\rho }_{n}}{{p}_{n}}+p_{n}^{c}$. Evidently, the third constraint of problem \eqref{eq5} is met when the one of problem \eqref{eq6} is met.
\par
Similar to the operation in work \cite{7}, the problem \eqref{eq6} can be transformed into a tractable form according to the following theorem.
\par
\noindent
\textbf{Theorem 1: }If $\left( \boldsymbol{\bar{x}},\boldsymbol{\bar{\omega }} \right)$ is the solution of problem \eqref{eq6}, then there exist $\boldsymbol{\bar{\lambda }}$, such that $\boldsymbol{\bar{x}}$ satisfies the Karush-Kuhn-Tucker (KKT) conditions of the following problem for $\boldsymbol{\lambda }=\boldsymbol{\bar{\lambda }}$ and $\boldsymbol{\omega}=\boldsymbol{\bar{\omega}}$.
\begin{equation}\label{eq7}
\begin{split}
  \underset{\boldsymbol{x}}{\mathop{\max }}\,&\ G\left( \boldsymbol{x} \right)=\sum\limits_{n\in \mathcal{N}}{\sum\limits_{k\in \mathcal{K}}{{{x}_{nk}}\left\{ {{\omega }_{nk}}-{{\alpha }_{n}}\sum\limits_{i\in \mathcal{K}}{{{\lambda }_{ni}}{{\omega }_{ni}}} \right\}}} \\
\text{s.t. }& \sum\limits_{n\in \mathcal{N}}{{{x}_{nk}}}=1,\forall k\in \mathcal{K}, \\
 &\sum\limits_{n\in \mathcal{N}}{{{x}_{nk}}{{\text{SINR}}_{nk}}}\ge {{\tau }_{k}},\forall k\in \mathcal{K}, \\
 &{{x}_{nk}}\in \left\{ 0,1 \right\},\forall n\in \mathcal{N},\forall k\in \mathcal{K}, \\
\end{split}
\end{equation}
In addition, $\boldsymbol{\bar{x}}$ also satisfies the following system equations for $\boldsymbol{\lambda }=\boldsymbol{\bar{\lambda }}$ and $\boldsymbol{\omega}=\boldsymbol{\bar{\omega}}$.
\begin{equation}\label{eq8}
{{\lambda }_{nk}}=\frac{{{x}_{nk}}}{{{\alpha }_{n}}\left( 1+\sum\nolimits_{i\in \mathcal{K}}{{{x}_{ni}}} \right)},\forall n\in \mathcal{N},\forall k\in \mathcal{K},
\end{equation}
\begin{equation}\label{eq9}
{{\omega }_{nk}}=\frac{{{r}_{nk}}}{{{\alpha }_{n}}\left( 1+\sum\nolimits_{i\in \mathcal{K}}{{{x}_{ni}}} \right)},\forall n\in \mathcal{N},\forall k\in \mathcal{K}.
\end{equation}
On the contrary, if $\boldsymbol{\bar{x}}$ is a solution of problem \eqref{eq7} and satisfies the equations \eqref{eq8} and \eqref{eq9} for $\boldsymbol{\lambda }=\boldsymbol{\bar{\lambda }}$ and $\boldsymbol{\omega}=\boldsymbol{\bar{\omega}}$, $\left( \boldsymbol{\bar{x}},\boldsymbol{\bar{\omega }} \right)$ also satisfies the KKT conditions of problem \eqref{eq6}.
\par
\emph{Proof: }As for the third constraint of problem \eqref{eq6}, we Introduce Lagrange multipliers $\boldsymbol{\lambda }=\left\{ {{\lambda }_{nk}},\forall n\in \mathcal{N},\forall k\in \mathcal{K} \right\}$. Then, the Lagrange function with respect to this constraint can be written as
\begin{equation}\label{eq10}
\begin{split}
\mathcal{L}\left( \boldsymbol{x},\boldsymbol{\lambda } \right)=&\sum\limits_{n\in \mathcal{N}}{\sum\limits_{k\in \mathcal{K}}{{{\lambda }_{nk}}\left\{ {{r}_{nk}}-{{\alpha }_{n}}{{\omega }_{nk}}\left( 1+\sum\limits_{i\in \mathcal{K}}{{{x}_{ni}}} \right) \right\}}}\\
&+\sum\limits_{n\in \mathcal{N}}{\sum\limits_{k\in \mathcal{K}}{{{x}_{nk}}{{\omega }_{nk}}}}.\\
\end{split}
\end{equation}
Since $\left( \boldsymbol{\bar{x}},\boldsymbol{\bar{\omega }} \right)$ is the solution of problem \eqref{eq6}, there exist $\boldsymbol{\bar{\lambda}}$ for any $n$ and $k$ such that partial KKT conditions of problem \eqref{eq6} are listed as follows
\begin{equation}\label{eq11}
\frac{\partial \mathcal{L}}{\partial {{\omega }_{nk}}}={{\bar{x}}_{nk}}-{{\alpha }_{n}}{{\bar{\lambda }}_{nk}}\left( 1+\sum\limits_{i\in \mathcal{K}}{{{{\bar{x}}}_{ni}}} \right)=0,
\end{equation}
\begin{equation}\label{eq12}
{{\bar{\lambda }}_{nk}}\frac{\partial \mathcal{L}}{\partial {{\lambda }_{nk}}}={{\bar{\lambda }}_{nk}}\left\{ {{r}_{nk}}-{{\alpha }_{n}}{{{\bar{\omega }}}_{nk}}\left( 1+\sum\limits_{i\in \mathcal{K}}{{{{\bar{x}}}_{ni}}} \right) \right\}=0.
\end{equation}
Evidently, the condition $1+\sum\nolimits_{i\in \mathcal{K}}{{{{\bar{x}}}_{ni}}}>0$ is met for any $n$. Thus, we can easily obtain the following results according to the equations \eqref{eq11} and \eqref{eq12}.
\begin{equation}\label{eq13}
{{\bar{\lambda }}_{nk}}=\frac{{{{\bar{x}}}_{nk}}}{{{\alpha }_{n}}\left( 1+\sum\nolimits_{i\in \mathcal{K}}{{{{\bar{x}}}_{ni}}} \right)},\forall n\in \mathcal{N},\forall k\in \mathcal{K},
\end{equation}
\begin{equation}\label{eq14}
{{\bar{\omega }}_{nk}}=\frac{{{r}_{nk}}}{{{\alpha }_{n}}\left( 1+\sum\nolimits_{i\in \mathcal{K}}{{{{\bar{x}}}_{ni}}} \right)},\forall n\in \mathcal{N},\forall k\in \mathcal{K}.
\end{equation}
It is easy to know that the equations \eqref{eq11} and \eqref{eq12}) are also the KKT conditions of the following problem for $\boldsymbol{\lambda}=\boldsymbol{\bar{\lambda}}$ and $\boldsymbol{\omega}=\boldsymbol{\bar{\omega}}$.
\begin{equation}\label{eq15}
\begin{split}
 \underset{\boldsymbol{x}}{\mathop{\max }}\,&\ H\left( \boldsymbol{x} \right)=\mathcal{L}\left( \boldsymbol{x},\boldsymbol{\lambda } \right) \\
\text{s.t. }&\sum\limits_{n\in \mathcal{N}}{{{x}_{nk}}}=1,\forall k\in \mathcal{K}, \\
 &\sum\limits_{n\in \mathcal{N}}{{{x}_{nk}}{{\text{SINR}}_{nk}}}\ge {{\tau }_{k}},\forall k\in \mathcal{K}, \\
 &{{x}_{nk}}\in \left\{ 0,1 \right\},\forall n\in \mathcal{N},\forall k\in \mathcal{K}. \\
\end{split}
\end{equation}
\par
Since the problem \eqref{eq15} is an optimization problem with respective to $\boldsymbol{x}$, it can be easily simplified into the problem \eqref{eq7}. Therefore, the first conclusion of Theorem 1 holds. According to the similar procedure, the contrary conclusion can also be easily proved.
\par
The mentioned-above theorem shows that the feasible solution of problem \eqref{eq6} can be found among all solutions of problem \eqref{eq7}, which satisfies the equations \eqref{eq8} and \eqref{eq9}. In addition, such a solution is also a global solution of problem \eqref{eq6} if it is unique \cite{7}.
\par
We first introduce Lagrange multipliers $\boldsymbol{\mu }=\left\{ {{\mu }_{nk}},\forall n\in \mathcal{N},\forall k\in \mathcal{K} \right\}$ for the second constraint of problem \eqref{eq7}, and then obtain a Lagrange function with respect to this constraint as follows:
\begin{equation}\label{eq16}
\begin{split}
\mathcal{L}\left( \boldsymbol{x},\boldsymbol{\mu } \right)=&\sum\limits_{n\in \mathcal{N}}{\sum\limits_{k\in \mathcal{K}}{{{x}_{nk}}\left\{ {{\omega }_{nk}}-{{\alpha }_{n}}\sum\limits_{i\in \mathcal{K}}{{{\lambda }_{ni}}{{\omega }_{ni}}} \right\}}}\\
&+\sum\limits_{k\in \mathcal{K}}{{{\mu }_{k}}\left( \sum\limits_{n\in \mathcal{N}}{{{x}_{nk}}{{\text{SINR}}_{nk}}}-{{\tau }_{k}} \right)}.\\
\end{split}
\end{equation}
Then, the dual function can be written as
\begin{equation}\label{eq17}
I\left( \boldsymbol{\mu } \right)=\left\{ \begin{split}
  \underset{\boldsymbol{x}}{\mathop{\max }}\,&\ \mathcal{L}\left( \boldsymbol{x},\boldsymbol{\mu } \right) \\
\text{s.t. }&\sum\limits_{n\in \mathcal{N}}{{{x}_{nk}}}=1,\forall k\in \mathcal{K}, \\
 &{{x}_{nk}}\in \left\{ 0,1 \right\},\forall n\in \mathcal{N},\forall k\in \mathcal{K}, \\
\end{split} \right.
\end{equation}
and the dual problem of problem \eqref{eq7} is
\begin{equation}\label{eq18}
\underset{\boldsymbol{\mu }\succcurlyeq \boldsymbol{0}}{\mathop{\min }}\,\ I\left( \boldsymbol{\mu } \right).
\end{equation}
where $\boldsymbol{a}\succcurlyeq \boldsymbol{b}$ if ${{a}_{nk}}\ge {{b}_{nk}}$ for any $n$ and $k$.
\par
When the dual optimal solution ${{\boldsymbol{\mu }}^{*}}$ , the problem \eqref{eq17} can be easily solved according to the following rule:
\begin{equation}\label{eq19}
{{n}^{*}}=\arg \underset{n\in \mathcal{N}}{\mathop{\max }}\,\left\{ {{\omega }_{nk}}+{{\mu }_{k}}{{\text{SINR}}_{nk}}-{{\alpha }_{n}}\sum\limits_{i\in \mathcal{K}}{{{\lambda }_{ni}}{{\omega }_{ni}}} \right\},\forall k\in \mathcal{K},
\end{equation}
The rule \eqref{eq19} shows that any user $k$ selects some BS ${{n}^{*}}$ to maximize its utility, i.e., the association objective of rule \eqref{eq19}. Next, the solution of outer problem \eqref{eq19} can be searched using gradient descent method, and given by
\begin{equation}\label{eq20}
\mu _{k}^{t+1}={{\left[ \mu _{k}^{t}-{{\xi }}\left( \sum\limits_{n\in \mathcal{N}}{x_{nk}^{t}\text{SIN}{{\text{R}}_{nk}}}-{{\tau }_{k}} \right) \right]}^{+}},\forall k\in \mathcal{K},
\end{equation}
where ${{\left[ z \right]}^{+}}=\max \left\{ z,0 \right\}$; ${{\xi }}$ represents a small enough stepsize.
\par
Based on the mentioned-above analyses, we can easily give a two-layer iterative algorithm to solve the problem \eqref{eq6}, which can be described in Algorithm 1. In this algorithm, the $\boldsymbol{\lambda }$ and $\boldsymbol{\omega }$ are updated using Newton-like method in the outer layer and $\boldsymbol{x}$ is decided by employing the rule \eqref{eq19} in the inner layer. In addition, we also give the definitions of some functions for any $n$ and $k$ as follows:
\begin{equation}\label{eq21}
{{\phi }_{nk}}\left( {{\lambda }_{nk}} \right)={{\alpha }_{n}}{{\lambda }_{nk}}\left( 1+\sum\limits_{i\in \mathcal{K}}{{{x}_{ni}}} \right)-{{x}_{nk}},
\end{equation}
\begin{equation}\label{eq22}
{{\varphi }_{nk}}\left( {{\omega }_{nk}} \right)={{\alpha }_{n}}{{\omega }_{nk}}\left( 1+\sum\limits_{i\in \mathcal{K}}{{{x}_{ni}}} \right)-{{r}_{nk}},
\end{equation}
\begin{equation}\label{eq23}
{{\chi }_{nk}}=\frac{1}{{{\alpha }_{n}}\left( 1+\sum\nolimits_{i\in \mathcal{K}}{{{x}_{ni}}} \right)}.
\end{equation}

\begin{table}[]
\centering
\begin{tabular}{ll}
\toprule[1pt]
\textbf{Algorithm 1: Energy-Efficient Association (EEA)} \\ \midrule[0.5pt]
1: \textbf{Initialization:} Let ${t_1}=1$, ${t_2}=1$, $\xi \in (0,1)$, $\varepsilon \in (0,1)$; initialize \\
\ \ \ ${{\boldsymbol{x}}^{{t}_{2}}}$, ${{\boldsymbol{\lambda}}^{{t}_{1}}}$,  ${{\boldsymbol{\mu}}^{{t}_{2}}}$ and ${{\boldsymbol{\omega}}^{{t}_{1}}}$. \\
2: \textbf{Repeat (Outer Loop)}\\
3: \ \ \ \ \textbf{Repeat (Inner Loop)}\\
4: \ \ \ \ \ \ \ \ Any user selects some BS according to the rule \eqref{eq19}.\\
5: \ \ \ \ \ \ \ \ Update multiplier $\mu _{k}^{{{t}_{2}}+1}$ using \eqref{eq20} for any $k$.\\
6: \ \ \ \ \ \ \ \ Normalize $\mu _{k}^{{{t}_{2}}+1}$, i.e.,  $\mu _{k}^{{{t}_{2}}+1}={\mu _{k}^{{{t}_{2}}+1}}/{{\sum\nolimits_{k\in \mathcal{K}}{\mu _{k}^{{{t}_{2}}+1}}}}$.\\
7: \ \ \ \ \ \ \ \ ${{t}_{2}}={{t}_{2}}+1$.\\
8: \ \ \ \ \textbf{Until} $G\left( \boldsymbol{x}\right)$ converges or ${{t}_{2}}={{T}_{2}}$.\\
9: \ \ \ \ If the following conditions are satisifed, then stop the algorithm. \\
\ \ \ \ \ \ \ Otherwise, go to step 5.\\
\ \ \ \ \ \ \ \ \ \ \ \ ${{\alpha }_{n}}\lambda _{nk}^{{t}_{1}}\left( 1+\sum\limits_{i\in \mathcal{K}}{x_{ni}^{{t}_{1}}} \right)-x_{nk}^{{t}_{1}}=0,\forall n\in \mathcal{N},\forall k\in \mathcal{K},$\\
\ \ \ \ \ \ \ \ \ \ \ \ ${{\alpha }_{n}}\omega _{nk}^{{t}_{1}}\left( 1+\sum\limits_{i\in \mathcal{K}}{x_{ni}^{{t}_{1}}} \right)-{{r}_{nk}}=0,\forall n\in \mathcal{N},\forall k\in \mathcal{K},$\\
10: \ \ \ \ Find the smallest $m$ among $m\in \left\{ 0,1,2,\cdots  \right\}$ satisfying\\
\ \ \ \ \ \ \ \ \ \ \ \ $\sum\limits_{n\in \mathcal{N}}{\sum\limits_{k\in \mathcal{K}}{{{\left| {{\phi }_{nk}}\left( \lambda _{nk}^{{t}_{1}}-{{\xi }^{m}}{{\chi }_{n}}{{\phi }_{nk}}\left( \lambda _{nk}^{{t}_{1}} \right) \right) \right|}^{2}}}}$ \\
\ \ \ \ \ \ \ \ \ \ \ \ \ \ \ +$\sum\limits_{n\in \mathcal{N}}{\sum\limits_{k\in \mathcal{K}}{{{\left| {{\varphi }_{nk}}\left( \omega _{nk}^{{t}_{1}}-{{\xi }^{m}}{{\chi }_{n}}{{\varphi }_{nk}}\left( \omega _{nk}^{{t}_{1}} \right) \right) \right|}^{2}}}}$ \\
\ \ \ \ \ \ \ \ \ \ \ \ $\le \left( 1-\varepsilon {{\xi }^{m}} \right)\sum\limits_{n\in \mathcal{N}}{\sum\limits_{k\in \mathcal{K}}{\left\{ {{\left| {{\phi }_{nk}}\left( \lambda _{nk}^{{t}_{1}} \right) \right|}^{2}}+{{\left| {{\varphi }_{nk}}\left( \omega _{nk}^{{t}_{1}} \right) \right|}^{2}} \right\}}}$ \\
11: \ \ \ \ Update $\boldsymbol{\lambda }$ and $\boldsymbol{\omega }$ according Newton-like method:\\
\ \ \ \ \ \ \ \ \ \ \ \ $\lambda _{nk}^{{{t}_{1}}+1}=\lambda _{nk}^{{t}_{1}}-{{\xi }^{m}}{{\chi }_{n}}{{\phi }_{nk}}\left( \lambda _{nk}^{{t}_{1}} \right),\forall n\in \mathcal{N},\forall k\in \mathcal{K},$ \\[+3pt]
\ \ \ \ \ \ \ \ \ \ \ \ $\omega _{nk}^{{{t}_{1}}+1}=\omega _{nk}^{{t}_{1}}-{{\xi }^{m}}{{\chi }_{n}}{{\varphi }_{nk}}\left( \omega _{nk}^{{t}_{1}} \right),\forall n\in \mathcal{N},\forall k\in \mathcal{K},$ \\[+3pt]
12: \ \ \ \ Normalize $\lambda _{nk}^{{{t}_{1}}+1}$, i.e.,  $\lambda _{nk}^{{{t}_{1}}+1}={\lambda _{nk}^{{{t}_{1}}+1}}/{\sum\nolimits_{n\in \mathcal{N}}{\sum\nolimits_{k\in \mathcal{K}}{\lambda _{nk}^{{{t}_{1}}+1}}}}$.\\
13: \ \ \ \ ${{t}_{1}}={{t}_{1}}+1$.\\
14: \textbf{Until} $F\left( \boldsymbol{x},\boldsymbol{\omega }\right)$ converges or ${{t}_{1}}={{T}_{1}}$.  \\ \bottomrule[0.5pt]
\end{tabular}
\label{alg1}
\end{table}
\par
In Algorithm 1, ${{t}_{1}}$ and ${{t}_{2}}$ are the iteration indices of outer and inner loops, ${{T}_{1}}$ and ${{T}_{2}}$ represent the maximal numbers of iterations of outer and inner loops, and the steps 6 and 12 normalize multipliers to ensure that the Lagrangian functions \eqref{eq10} and \eqref{eq16} are bounded. To prove the convergence of Algorithm 1, we give the following lemma.
\par
\noindent
\textbf{Lemma 1:} Algorithm 1 is convergent.
\par
 \emph{Proof: } In the outer layer loop of Algorithm 1, $\boldsymbol{\lambda }$ and $\boldsymbol{\omega }$ are updated using Newton-like method with a linear convergence rate. When ${{\xi }^{m}}=1$, the update of $\boldsymbol{\lambda }$ and $\boldsymbol{\omega }$ in the step 10 reduces to Newton method with a quadratic convergence rate. In addition, the convergence of outer loop of Algorithm 1 can be also proven by using a similar method used in work \cite{7}. Next, we show the convergence of inner loop of Algorithm 1. Since the problem \eqref{eq7} is a linear optimization problem with respective to $\boldsymbol{x }$, the inner loop will converge to its dual optimum in a bounded region. When the outer layer loop and inner layer loop are convergent, the whole algorithm should be convergent.
 \par
 Next, we will give some complexity analyses for the proposed algorithm. In Algorithm 1, the computation complexity of inner layer loop mainly depends on the BS selection i.e., step 4. Evidently, the inner layer loop has a computation complexity of $\mathcal{O}\left( {T_2}NK \right)$. Since the term $\sum\nolimits_{i\in \mathcal{K}}{x_{ni}}$ can be calculated before performing the step 9 of outer layer loop, the step 9 has a complexity of $\mathcal{O}\left( NK \right)$. Considering that the step 10 of outer layer loop needs to find the smallest $m$, we can deduce that such a step should have a complexity of $\mathcal{O}\left( \left( m+1\right)NK \right)$. As for other steps of outer layer loop, we can easily know that they have a complexity of $\mathcal{O}\left( NK \right)$. Thus, the computation complexity of outer layer loop is $\mathcal{O}\left( \left( m+1 \right){{T}_{1}}NK \right)$, where $m$ often takes a relatively small integer number. In general, the computation complexity of Algorithm 1 is the maximum between $\mathcal{O}\left( {T_1}{T_2}NK \right)$ and $\mathcal{O}\left( \left( m+1 \right){{T}_{1}}NK \right)$.
\section{Numerical Simulation}
In HCNs, the MBSs are fixed and deployed into a traditional cellular network, while the PBSs and users are distributed at each macrocell in a relatively random manner. We assume that the inter-site distance between any two MBSs is 1 km, the maximal transmit powers of MBS and PBS are 46 dBm and 30 dBm respectively, the circuit power of one PBS is 13.6 W, the power amplifier coefficients of MBS and PBS are 4 and 2 respectively, the noise power spectral density is -174 dBm/Hz and the system bandwidth is 10 MHz. In HCNs, we adopt the pathloss models $l_{nk}=128.1+37.6\log 10\left( d_{nk} \right)$ and $l_{nk}=140.7+36.7\log 10\left( d_{nk} \right)$ for MBS and PBS respectively, where $d_{nk}$ is the distance (in km) between user $k$ and BS $n$. Meanwhile, we also consider a log-normal shadowing with a standard deviation of 8 dB in the propagation environment. In addition, we need to set the coefficients mentioned in the formula \eqref{eq4}. Specially, we let ${{C}_{00}}=4$, ${{C}_{10}}=4.8$, ${{C}_{20}}=0$ ,${{C}_{30}}=2.08\times {{10}^{-8}}$, ${{C}_{01}}=1$, ${{C}_{11}}=9.5\times {{10}^{-8}}$ and ${{C}_{21}}=6.25\times {{10}^{-8}}$.
\par
Next, we will investigate the association performance of proposed association (energy-efficient association, EEA), which includes average rate, average energy efficiency and the convergence of proposed algorithm. To highlight the effectiveness of scheme EAA, we introduce another energy-efficient association with user fairness (EEAUF) for comparison.

\begin{figure}[!t]
\centering
\centerline{\includegraphics[width=3.5in]{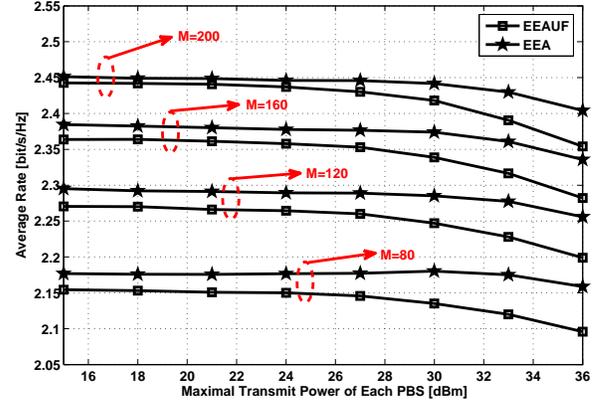}}
\caption{Impacts of maximal transmit power of each PBS on the average rates under different numbers of MBS antennas for different associations.}
\label{fig1}
\end{figure}
\par
Under different numbers of MBS antennas, Fig. \ref{fig1} shows the impacts of maximal transmit power of each PBS on the average rates for different associations, where the average rate represents the average of effective rates of all associated users. In general, the average rate decreases with the maximal allowed transmit power of each PBS. According to the system model, it is easy to find that there are two types of reasons for this trend. In the first type, the MBSs install large-scale antenna array but the PBSs just adopt signal antenna model. In the second type, the MBSs often have a higher transmit power than PBSs. These two points ensure that the signal strength of massive MIMO enabled MBSs is often far stronger than the one of PBSs, which results in a case that most users are attracted by MBSs and very few users can select PBSs. Although the increment of transmit power of each PBS can increase the effective rates of pico users (users associated with PBSs), it can also decrease the effective rates of macro users (users associated with MBSs) due to increased interference. Since the number of pico users is often far litter than the one of macro users, the average rate may decrease with maximal transmit power of each PBS. As illustrated in Fig. \ref{fig1}, the average rate increases with the number of MBS antennas. The reason for this is that the signal strength of MBSs can be increased by increasing the number of MBS antennas. Although it often results in a decrement of effective signal strength of PBSs due to the increased interference, the average rate may increase with it since the number of macro users is often far larger than the one of pico users. In addition, the scheme EEA has a higher average rate than scheme EEAUF since it doesn't guarantee user fairness.
\begin{figure}[!t]
\centering
\centerline{\includegraphics[width=3.5in]{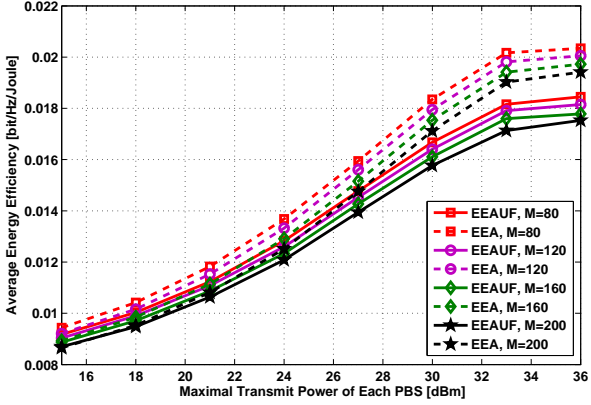}}
\caption{Impacts of maximal transmit power of each PBS on the average energy efficiencies under different numbers of MBS antennas for different associations.}
\label{fig2}
\end{figure}
\par
Under different numbers of MBS antennas, Fig. \ref{fig2} shows the impacts of maximal transmit power of each PBS on the energy efficiencies for different associations, where the average energy efficiency represents the average of energy efficiencies of all associated users. In general, the average energy efficiency increases with the maximal transmit power of each PBS, but it may decrease with this parameter when the transmit power of each PBS is high enough. As we know, the increased transmit power can improve the experience of pico users in low power domain, but it then decreases the experience because of stronger and stronger intra-tier interference in high power domain. Since the massive MIMO enabled MBSs often have a far higher transmit power than PBSs and the transmit power of the former is often far larger than its effective rates, the change of effective rates of MBSs may not have an evident influence on the energy efficiency of macro users. Therefore, the change of transmit power of PBSs mainly affect the energy efficiency of pico users. Evidently, the average energy efficiency may increase with the transmit power of each PBS due to the increased signal strength, but it may decrease with this power due to the increased interference and increased transmit power. As shown in the formula \eqref{eq4}, the power consumption scales with the number of MBS antennas. Thus, the average energy efficiency should decrease with the number of MBS antennas. In addition, the scheme EEA has a higher average energy efficiency than scheme EEAUF because it doesn't guarantee user fairness.
\begin{figure}[!t]
\centering
\centerline{\includegraphics[width=3.5in]{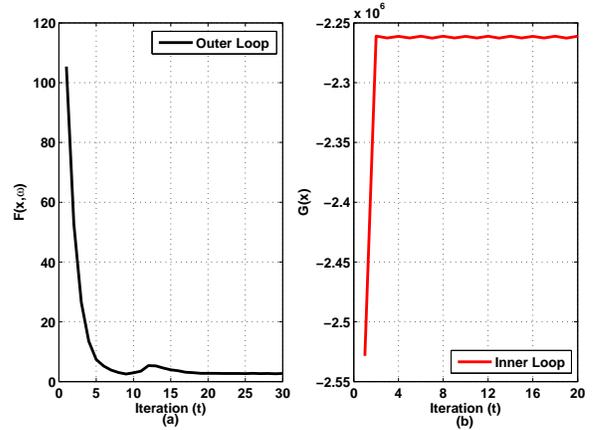}}
\caption{The convergence of proposed algorithm.}
\label{fig3}
\end{figure}
\par
Fig. \ref{fig3} shows the convergence of proposed algorithm, where $t$ represents iteration index. Specially, the Fig. \ref{fig3} (a) shows the convergence of outer (layer) loop in Algorithm 1, and Fig. \ref{fig3} (b) shows the convergence of inner (layer) loop in Algorithm 1. Through a direct observation, it is easy to find that these loops have relatively fast convergence rates and the inner loop may have volatility convergence.
\section{CONCLUSION}
In this paper, we design an energy-efficient association scheme from a novel perspective, and formulate it as a sum energy efficiency maximization problem under users' SINR constraints. Considering that the formulated problem is in a sum-of-ratio form and hard to tackle, we first need to transform it into a tractable form through a parametric programming. Then, we try to develop a two-layer iterative algorithm with guaranteed convergence to achieve its solution. After that, we give some convergence and complexity analyses for the proposed algorithm. Numerical results show that the proposed scheme has a higher system throughput and a higher network energy efficiency than the existing scheme. In addition, we also investigate the impacts of the number of MBS antennas and the transmit power of each PBS on these association performances.




%
%
%
\bibliographystyle{IEEEtran}
\bibliography{reference}

\end{document}